\documentclass[%
 reprint,
showpacs,preprintnumbers,
showkeys,
 amsmath,amssymb,
 aps,
 prl,
]{revtex4-1}

\usepackage[english]{babel}
\usepackage[utf8]{inputenc}
\usepackage{graphicx}
\usepackage{bm}
\usepackage{hyperref}
\usepackage{xcolor}
\usepackage{amsmath}
\usepackage{amsfonts}
\usepackage{amssymb}
\usepackage{amsthm}
\usepackage{dsfont}

\hypersetup{
    unicode=false,                 
    pdftoolbar=true,               
    pdfmenubar=true,               
    pdffitwindow=false,            
    pdfstartview={FitH},           
    pdftitle={},                   
    pdfauthor={Nicolai Lang},      
    pdfsubject={Quantum physics},  
    pdfcreator={Nicolai Lang},     
    pdfproducer={Nicolai Lang},    
    pdfkeywords={},                
    pdfnewwindow=true,             
    colorlinks=true,               
    linkcolor=cyan,             
    citecolor=magenta,                
    filecolor=magenta,             
    urlcolor=blue                  
}

\def\bra#1{\mathinner{\langle{#1}|}}

\def\Bra#1{\left<#1\right|}
\def\Ket#1{\left|#1\right>}

\renewcommand{\vec}[1]{\mathbf{#1}}
\renewcommand\arraystretch{2.0}
\newcommand{\tr}[1]{\operatorname{Tr}\left[#1\right]}

\newcommand{\Tr}[2]{\operatorname{Tr}_{#2}\left[#1\right]}

\newcommand{\com}[2]{\left[#1,#2\right]}
\newcommand{\acom}[2]{\left\{#1,#2\right\}}

\graphicspath{{./}}

\begin{document}


\title{Exploring quantum phases by driven dissipation}

\author{Nicolai Lang}
\email{nicolai@itp3.uni-stuttgart.de}
\author{Hans Peter Büchler}
\email{buechler@theo3.physik.uni-stuttgart.de}
\affiliation{Institute for Theoretical Physics III, 
  University of Stuttgart, 70550 Stuttgart, Germany}

\date{\today}


\begin{abstract}

  Ever since the insight spreaded that tailored dissipation can be employed to control quantum systems and drive
  them towards pure states, the field of non-equilibrium quantum mechanics gained remarkable momentum.
  So far research focussed on emergent phenomena caused by the interplay and competition of unitary Hamiltonian and dissipative Markovian dynamics.
  In this manuscript we zero in on a so far rather understudied aspect of open quantum systems and non-equilibrium physics,
  namely the utilization of \textit{purely dissipative} couplings to explore pure quantum phases and non-equilibrium phase transitions.
  To illustrate this concept, we introduce and scrutinize purely dissipative counterparts of (1) the paradigmatic transverse field Ising model
  and (2) the considerably more complex $\mathbb{Z}_2$ lattice gauge theory with coupled matter field. We show that, in mean field
  approximation, the non-equilibrium phase diagrams parallel the (thermal) phase diagrams of the Hamiltonian ``blue print'' theories qualitatively.

\end{abstract}

\pacs{}


\maketitle

Both dissipative quantum computation \cite{Verstraete2009,Pastawski2011} and state preparation \cite{Kraus2008,2_weimer_RQS_A,Ticozzi2013} are based on the description 
of the quantum system in terms of a Lindblad master equation. 
Both require the existence of a unique and pure state as non-equilibrium steady state (NESS), which is a dark state of the dissipative coupling between system and bath, 
i.e., the state does not interact with the open reservoir. 
Especially the existence and uniqueness of the desired pure steady state is in general a highly non-trivial task, and requires often a careful and sophisticated design 
of the coupling between system and bath. 
For example, it has been proven that any graph state can be prepared efficiently by dissipation \cite{Verstraete2009,Kraus2008}; 
the latter being a ressource for dissipative quantum computation. 
First \textit{experimental} proofs of principle of these ideas have been furnished quite recently 
with trapped ions \cite{Barreiro2011,Schindler2013}.
In such experimental setups the implementation of theoretically well-designed couplings will be error-prone and, in general, lead to a 
\textit{mixed} steady state. It is then a crucial question whether this non-equilibrium steady state is ``close enough'' to the desired pure dark state
and still features the desired properties. First steps into this direction have been taken by analyzing the appearance of non-equilibrium phase transitions 
due to competing coherent and dissipative dynamics \cite{Diehl2008,Prosen2008a,Diehl2010a,Eisert2010a,Tomadin2011,Foss-Feig2012,Ates2012,Kessler2012a,Shirai2012,Lesanovsky2013,Banchi2013}. 

In this manuscript, we study this question in a paradicmatic setup, where competing dissipative terms drive the system towards well-known pure quantum phases 
and, as a consequence, give rise to a non-equilibrium phase transition connecting them. 
The central idea is to start with two different types of dissipative terms: the first one drives the system into a unique and pure non-equilibrium steady state, 
whereas the second type of dissipative coupling prefers steady states exhibiting true long-range order. We then analyze the non-equilibrium phase diagram 
depending on the relative coupling strength of the two dissipative baths. 
This analysis follows a mean field treatment of the dissipative dynamics --- which is valid in high dimensions. 
We derive the properties of the phase transition as well as its critical exponents, and compare its behavior with the well-established thermal phase transition 
of the analog Hamiltonian theory. We argue that such purely dissipative quantum simulations can pave the way for the robust exploration of phase diagrams 
of complex quantum systems that are notoriously hard to tackle analytically. 
Building on these observations, we expand our concept and present a dissipative quantum simulation of the $\mathbb{Z}_2$ lattice gauge theory with coupled matter field.

\begin{figure}[t]
  \includegraphics[width=1.0\linewidth]{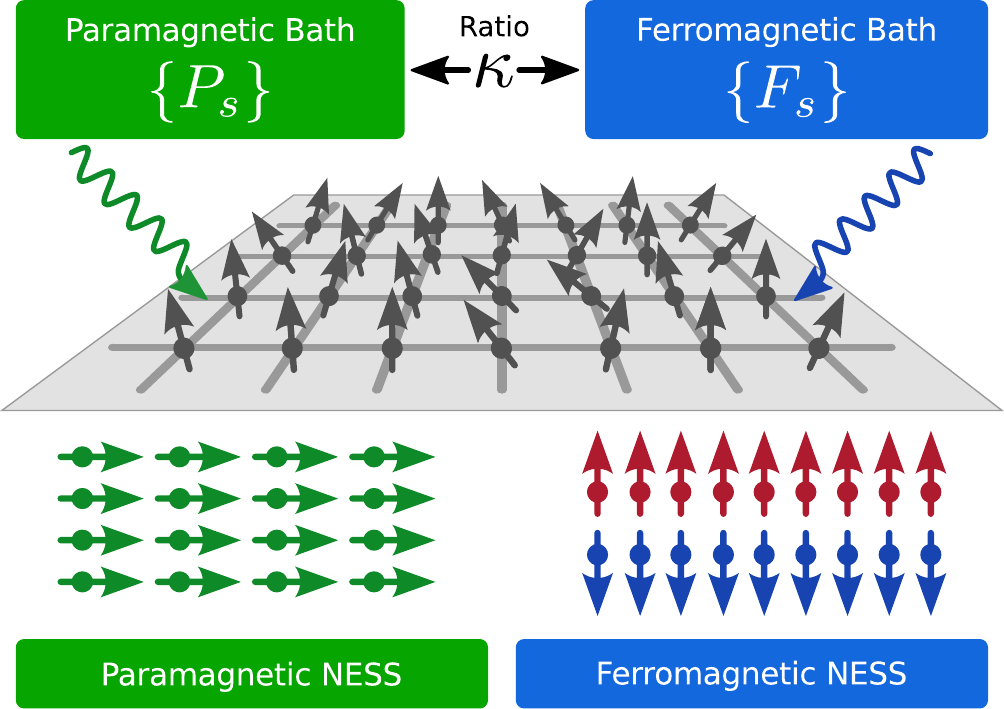}
  \caption{
    Schematic setup.
    We consider a $D$-dimensional rectangular lattice with spins attached to the sites.
    The system is homogeneously coupled to \textit{two} tailored markovian baths with relative
    coupling strength $\kappa$.
    The $\{P_s\}$ ($\{F_s\}$) jump operators drive the system towards the paramagnetic (ferromagnetic)
    ground states of the transverse field Ising model.
    There is \textit{no} unitary dynamics involved.
  }
  \label{fig:TIM_setup}
\end{figure}

We start with a description of the time evolution of a generic quantum system coupled to a Markovian bath.  
Throughout our manuscript we are interested in a purely dissipative dynamics governed by the Lindblad master equation 
\cite{Lindblad1976a}
\begin{equation}
 \label{eq:lindblad}
 \dot\rho=\sum_i\,\left[L_i\rho L_i^\dagger -\frac{1}{2}\left\{L_i^\dagger L_i,\rho\right\}\right]\equiv\mathcal{L}\rho
\end{equation}
with non-hermitian jump operators $\{L_i\}$ characterizing the microscopic actions of the bath(s). 
Here $\rho$ denotes the system density matrix and $\{\bullet,\bullet\}$ the anti-commutator.
$\mathcal{L}$ is termed \textit{Lindblad superoperator} and generates the semi-group of completely positive trace-preserving maps $\exp(\mathcal{L}t)$ ($t\geq 0$)
which describes the time evolution via $\rho(t)=\exp(\mathcal{L}t)\rho_0$.
Fixed points $\mathcal{L}\rho_\text{NESS}=0$ in the convex set of density matrices are usually refered to as \textit{non-equilibrium steady states};
the pure ones $\rho_\text{NESS}=\Ket{\Psi}\Bra{\Psi}$ for which $L_i\Ket{\Psi}=0$ holds for all jump operators $L_i$ are particularly interesting and
called \textit{dark states} \cite{Kraus2008}. The dynamics described by the Lindblad equation~(\ref{eq:lindblad}) is completely determined by the jump operators $\{L_i\}$,
the physical origin of which can be interpreted in various ways: From a \textit{microscopic} angle they can be
taken as the \textit{effective action} of a Hamiltonian environment by tracing out its unitary dynamics and using the Born-Markov approximation
(alongside additional assumptions) \cite{Plenio1998}. 
A different and more flexible point of view emerges in the field of digital quantum simulation \cite{1_feynman_SPWC,Lloyd1996,2_weimer_RQS_A} where the local jumps $L_i$
are realised explicitely by the simulator in terms of local, tailored interactions. As we are interested in a \textit{generic simulation} of quantum phases,
we shall take the latter point of view and omit any microscopic realisations of the contrived jump operators. To this end we point out that a scheme for the
microscopic simulation of arbitrary (local) jump operators was introduced in Ref.~\cite{2_weimer_RQS_A}.

\paragraph{A paradigmatic model.}

We start with a well-known model featuring a quantum phase transition: the transverse field Ising model (TIM)~\cite{Sachdev201105}.
The Hamiltonian for this paradigmatic theory on a $D$-dimensional (hyper-)cubic lattice with the spins located on sites $s\in\mathbb{S}$ reads
\begin{equation}
H_{\text{TIM}}=-J\sum_{\langle s,t\rangle}\sigma_s^z\sigma_t^z-h\sum_s\sigma_s^x,  
\end{equation}
where $J\geq 0$ determines the nearest-neighbour coupling strength and $h$ the transverse magnetic field. Here,
$\sigma_s^\mu$ ($\mu=x,y,z$) are the Pauli matrices that act on spin $s$.

The appearance of a quantum phase transition and the properties of the different phases are well understood in the two
limiting cases: for $h/J  \rightarrow \infty$  we recover the disordered 
ground state $\Ket{+}^\mathbb{S}$ which characterizes the \textit{paramagnetic phase}, whereas for $h/J\to 0$ the system reaches the \textit{ferromagnetic phase}
with the two-fold degenerate, symmetry-broken ground states $\Ket{\uparrow}^\mathbb{S}$ and $\Ket{\downarrow}^\mathbb{S}$.

%
%
%

These observations serve as a ``blue-print'' to construct a \textit{dissipative analogue} of the transverse field Ising model.
The main idea is to contrive \textit{two competing baths} such that the dark states of the individual
baths coincide with the ground states of the Hamiltonian theory in the above limiting cases. 
This concept allows us, first, to explore the quantum phases of the original Hamiltonian theory in a purely dissipative setup, and, second, to observe
a non-equilibrium counterpart of the symmetry-breaking quantum phase transition mentioned above. 
The jump operators for the \textit{dissipative transverse field Ising model} take the form
(an interpretation of their actions follows below),
\begin{subequations}
  \label{eq:tim_jops}
  \begin{eqnarray}
    P_s&=&\sqrt{\kappa}\,\sigma^z_s\left[\mathds{1}-\sigma^x_s\right],\quad\text{and}\\
    F_s&=&\sigma_s^x\left[\mathds{1}-\frac{1}{q}\sum_{t\in s}\sigma_{t}^z\sigma_s^z\right]
    \equiv\sigma_s^x\left[\mathds{1}-\sigma_{t\in s}^z\sigma_s^z\right],
  \end{eqnarray}
\end{subequations}
where $\kappa\geq 0$ is the relative coupling strength of the two baths (in analogy to the ratio  $h/J$ in the Hamiltonian theory).
Here we introduced the convenient notation $O_{t\in s}\equiv \frac{1}{|s|}\sum_{t\in s}O_t$,
where $\sum_{t\in s}$ denotes the sum over all sites $t$ adjacent to site $s$ and $|s|=q=2D$ denotes the number of nearest neighbours.
Please note that the complete dissipative process $\{L_i\}=\{P_s,F_s\}$ decomposes into two competing baths of relative strength $\kappa$, the \textit{paramagnetic} bath $\{P_s\}$ and the \textit{ferromagnetic} bath $\{F_s\}$, each of which acts translationally invariant on all sites $s$.
Clearly, the dissipative process $\{P_s,F_s\}$ inherits the global $\mathbb{Z}_2$-symmetry $U=\prod_s\sigma_s^x$ of the transverse field Ising model, 
namely $UL_sU^\dag=e^{i\alpha}L_s$, $\alpha\in [0,2\pi)$ for all $L_s=P_s,F_s$.
This setup is illustrated schematically in Fig.~\ref{fig:TIM_setup}.

The construction of the jump operators in~(\ref{eq:tim_jops}) follows the generic template
\begin{equation*}
L=\text{THEN}\cdot\text{IF}
\end{equation*}
where the $\text{IF}$-part ``checks'' whether some condition is met and the $\text{THEN}$-part
applies a conditioned action thereupon. For the paramagnetic jump operators $P_s$ this reads $\text{IF}=\mathds{1}-\sigma^x_s$
which probes whether the spin points along the magnetic field axis, and flips the spin otherwise via $\text{THEN}=\sigma^z_s$, hence 
driving the system towards the disordered ground state $\Ket{+}^\mathbb{S}$. 
The ferromagnetic jump operators $F_s$ count the number of antiparallel neighbours via $\text{IF}=\mathds{1}-1/q\sum_{t\in s}\sigma_{t}^z\sigma_s^z$
and condition thereby the spin flip $\text{THEN}=\sigma^x_s$, driving towards the completely correlated ground states 
$\alpha\Ket{\uparrow}^\mathbb{S}+\beta e^{i\phi}\Ket{\downarrow}^\mathbb{S}$, where $|\alpha|^2+|\beta|^2=1$ and $\phi\in [0,2\pi)$.

\begin{figure*}[t]
  \includegraphics[width=1.0\linewidth]{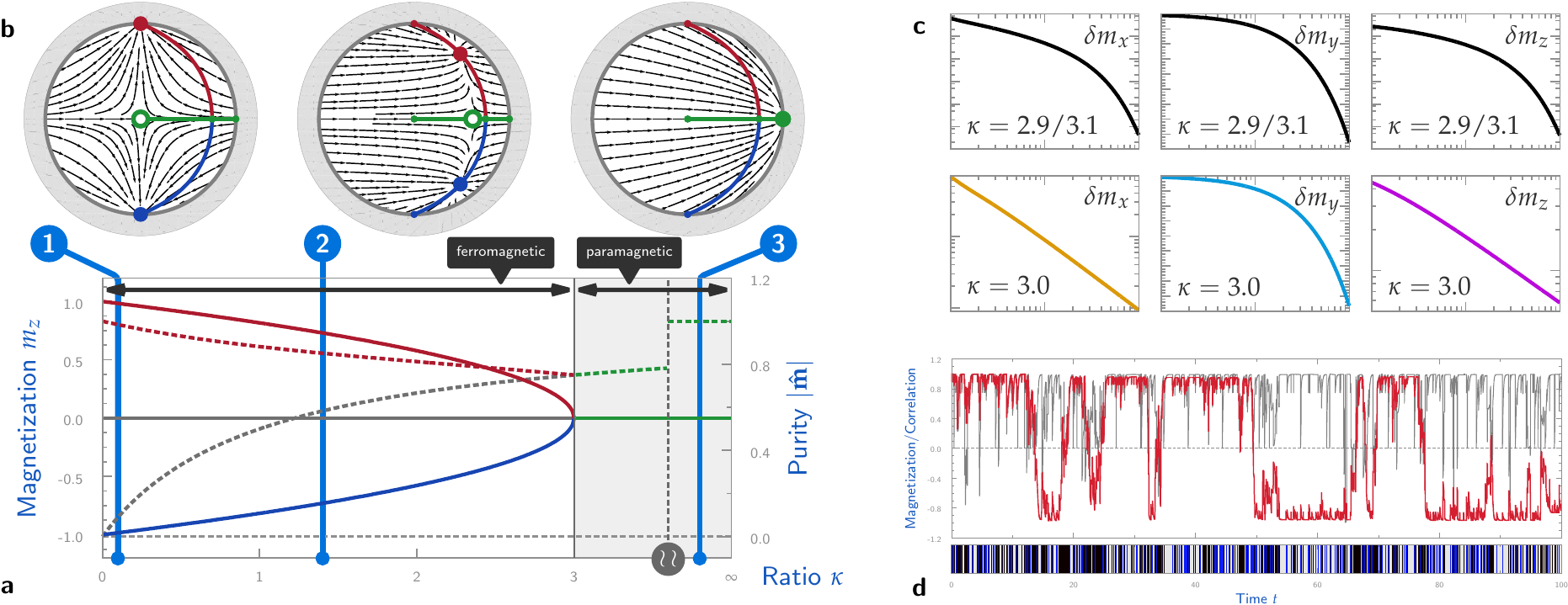}
  \caption{
    Results for the dissipative transverse field Ising model. (A) Mean field phase diagram. We show the magnetization $m_z$ (solid lines) for all steady states as a function of $\kappa$
    [red/blue: stable ferromagnetic; grey: unstable paramagnetic; green: stable paramagnetic]. The corresponding purities $|\hat{\vec m}|$ are illustrated by dashed lines with the same colours.
    (B) Dynamical mean field Lindblad flow $\vec F(\vec m)$ in the $m_x$-$m_z$-plane of the Bloch ball. 
    Stable (unstable) steady states are labeled by bullets (circles); their paths for $0\leq\kappa\leq\infty$ are highlighted. We illustrate the flow for three (1,2 and 3) different ratios $\kappa$ above and below the 
    critical ratio $\kappa_c=3$. (C) Relaxation of the Bloch vector $\vec m(t)=\delta\vec m(t)+\hat{\vec m}$ close to the steady state below, at and above the critical ratio. 
    The relaxation in $m_x$- and $m_z$-direction becomes polynomial at the phase transition. (D) Quantum jump trajectory of a $3\times 3$ instance with periodic boundary conditions. We show the average magnetization $1/9\langle\sum_s\sigma_s^z\rangle$ (red)
    and the correlation $\langle\sigma_1^z\sigma_2^z\rangle$ (gray) starting from a completely polarized state $\Ket{\uparrow}^{\otimes 9}$. The ferromagnetic (paramagnetic) jumps $F_s$ ($P_s$) are 
    encoded by blue (black) impulses in the lower part.
  }
  \label{fig:TIM_plot}
\end{figure*}


Along the lines of the Hamiltonian theory (where quantum phases are characterized by the ground state(s)),
we are interested in the \textit{non-equilibrium steady states} $\rho_\text{NESS}$ of the dissipative theory with $\mathcal{L}\rho_\text{NESS}=0$, 
which characterize the non-equilibrium phases.  It immediately follows from the design of the jump operators, that in the limit  $\kappa\to \infty$ the steady state 
is a unique dark state and coincides with the disordered pure state $\rho_\text{NESS}=\Ket{+}\bra{+}^\mathbb{S}$, 
whereas for $\kappa\to 0$ the steady states are determined by the two symmetry-broken dark states $\Ket{\uparrow}^\mathbb{S}$ and 
$\Ket{\downarrow}^\mathbb{S}$, as well as coherent and incoherent mixtures thereof.
%
In the latter case, all steady states exhibit long range order $\langle\sigma_i^z\sigma_j^z\rangle=1$ for $|i-j|\to\infty$ --- just as in the case of the Hamiltonian transverse
field Ising model.
%
Finally, for a finite bath ratio ($0<\kappa<\infty$) there are no dark states \footnote{This is easy to see since there is no common pure state in the 
kernels of all jump operators $P_s,F_s$.} and the system is driven towards a (unique, as simulations suggest)
mixed steady state.
It is therefore natural to ask whether there is a non-trivial dissipatively driven phase transition (in the thermodynamic limit) 
from a high-$\kappa$ disordered to a low-$\kappa$ ordered phase, which may be considered a non-equilibrium analogue of 
the transverse field Ising model phase transition.

\paragraph{Mean field theory.}
To tackle this question, we analyze the phase diagram of the driven dissipative transverse field Ising model within mean field theory, which will provide
 reliable results for large lattice dimensions $D$.
%
The basic procedure to derive an effective mean field description for Lindbladian theories is quite similar to the Hamiltonian counterpart~\cite{Tomadin2011}:
We start with the product ansatz $\rho=\bigotimes_s\rho_s$ for the density matrix ($\rho_s$ denotes a single site density matrix)
and insert it into the Lindblad equation~(\ref{eq:lindblad}), thereby neglecting all spin-spin correlations. Tracing out the whole system except one
spin (and assuming a homogeneous system) yields an effective Lindblad equation for a single spin
\begin{equation}
  \label{eq:mf_lindblad}
  \dot{\hat\rho}=\sum_{j=0}^3 \left[\,f_j\hat\rho f_j^\dagger -\frac{1}{2}\left\{f_j^\dagger f_j,\hat\rho \right\} \right],
\end{equation}
where we set $\hat\rho\equiv \rho_s$ to emphasize the homogeneity of the system 
(i.e. the dynamics of the whole system decouples into the same single-spin dynamics for each spin).
The ferromagnetic jump operators give rise to three effective mean field jump operators, namely
\begin{eqnarray*}
  f_1&=&\sigma^x\left[\mathds{1}-m_z\sigma^z\right]\,,\\
  f_2&=&1/\sqrt{2D}\,\sqrt{1-m_z^2}\,\sigma^y\,,\\
\text{and}\qquad f_3&=&1/\sqrt{2D}\,\sigma^z,
\end{eqnarray*}
whereas the paramagnetic jump operator is not affected by the approximation, that is, $f_0=\sqrt{\kappa}\,\sigma^z\left[\mathds{1}-\sigma^x\right]$.
Note that interacting jump operators (such as $F_s$) result in more than one mean field jump operator
(here $f_{1,2,3}$) which account for dephasing due to the adjacent jump operators of the same type.
The expectation values $m_k\equiv\langle\sigma^k\rangle=\tr{\hat\rho \sigma^k}$ ($k=x,y,z$) have to be determined 
\textit{self-consistently} and thus render the mean field master equation non-linear in the single-spin density 
matrix $\hat\rho=(\mathds{1}+\vec m\vec\sigma)/2$ with the Bloch vector $\vec m=(m_x,m_y,m_z)$
restricted to $|\vec{m}| \leq 1$. Here self-consistency is ensured by identification of the expectation values $\langle\sigma^k\rangle$ and the 
Bloch vector components $m_k$. 

It it convenient to rewrite the Lindblad equation~(\ref{eq:mf_lindblad}) in terms of a
dynamical system 
\begin{equation}
  \label{eq:mf_system}
  \partial_t\vec m=\vec F(\vec m)
\end{equation}
with the non-linear flow $\vec F\,:\,\mathbb{R}^3\rightarrow\mathbb{R}^3$.
The steady state Bloch vectors $\hat{\vec m}$ are then determined by $\vec F(\hat{\vec m})=0$ and
their stability (i.e. physical relevance) can be inferred from the negativity of the spectrum of the Jacobian matrix
$\mathrm{D}\vec F(\hat{\vec m})$.
For technical details we refer the reader to the methods section.

\paragraph{Results.}

The main results of the mean field theory are outlined in Fig.~\ref{fig:TIM_plot}~(A)-(C).
We find a second order phase transition for our purely dissipative replica of the transverse field Ising model,
see Fig.~\ref{fig:TIM_plot}~(A). For the critical
mean field ratio one obtains $\kappa_c=4(1-1/q)$ which depends on the coordination number $q=2D$ (see methods).
For $\kappa\geq\kappa_c$ there is a single (stable) fixed point of $\vec F$
as can be seen from the $m_x$-$m_z$ cross section of the Bloch ball.
Starting from the correct paramagnetic dark state $\Ket{+}$ for $\kappa=\infty$, see~(B3), 
the steady state becomes mixed for $0<\kappa<\infty$ but remains paramagnetic until at $\kappa=\kappa_c$
two additional \textit{ferromagnetic} fixed points emerge. In the ferromagnetic regime $0\leq\kappa<\kappa_c$,
see~(B2), the paramagnetic solution becomes unstable. The ferromagnetic solutions reach the
correct dark states $\Ket{\uparrow}$ and $\Ket{\downarrow}$ for $\kappa\to 0$, see (B1).
At the critical point we find the typical mean field exponent $\beta=1/2$, i.e.,  $|\hat m_z|=(1-\kappa/\kappa_c)^\beta$.

In addition, the Lindblad master equation~(\ref{eq:mf_system}) provides information on the \textit{dynamics} of the system
and the time scales required to reach the steady state. Here we find a non-equilibrium critical slowing down
close to the phase transition, Fig.~\ref{fig:TIM_plot}~(C): Whereas above and below $\kappa_c$ the system
is damped exponentially close to the steady state, this decay turns out to be algebraic in $m_x$- and $m_z$-direction at
the phase transition, that is, $\delta m_k(t)\propto t^{\eta_k}$ ($k=x,z$) for $|\delta m_k|\ll 1$ (or $t\to\infty$) with the exponents $\eta_x=-1$ and $\eta_z=-1/2$.
We point out that the algebraic relaxation in $m_z$-direction with $\eta_z=-1/2$ is an immediate consequence of a vanishing eigenvalue of the Jacobian matrix $D\vec F$
(for $\kappa\neq \kappa_c$ it is negative-definite). In contrast, the algebraic relaxation in $m_x$-direction with $\eta_x=-1$ results from the coupling of $m_z$ and $m_x$
in Eq.~(\ref{eq:mf_system}) and different relaxation rates in $m_x$- and $m_z$-direction.

These results parallel the well-known mean field theory for the transverse field Ising model at finite temperatures 
(since the steady state is mixed at the phase transition, see Fig.~\ref{fig:TIM_plot}~(A)).
Nevertheless, this is a \textit{non-equilibrium} phase transition connecting the two zero temperature \textit{quantum} phases
of the transverse field Ising model via a \textit{non-thermal} manifold of states.

\paragraph{Monte Carlo Simulation.} 


In order to demonstrate the competitive nature of the baths $\{P_s\}$ and $\{F_s\}$ --- which is a key ingredient for the non-equilibrium phase transition ---, 
we performed quantum trajectory Monte Carlo (QTMC) simulations on small setups \cite{Revi1992,Dum1992,Plenio1998}.
A typical quantum jump trajectory for a $3\times 3$ lattice with periodic boundary conditions is shown in Fig.~\ref{fig:TIM_plot}~(D).
The initial state was completely $m_z$-polarized, $\Ket{\Psi_0}=\Ket{\uparrow}^{\otimes 9}$, 
and the bath ratio $\sqrt{\kappa}=1/3$ deep in the ferromagnetic regime.
We show the average polarization $1/9\langle\sum_s\sigma_s^z\rangle$ (red) and the nearest-neighbour correlation $\langle\sigma_1^z\sigma_2^z\rangle$ (gray). 
The ferromagnetic (paramagnetic) jumps $F_s$ ($P_s$) are encoded by blue (black) impulses below the plot.\\
The finite correlations --- combated by paramagnetic jumps --- indicate the emergence of local order due to the ferromagnetic driving. 
As a finite-size artifact, we observe a dynamically bistable behaviour of the polarization 
due to the competition of (dominant) ferromagnetic jumps stabilizing the plateaus and weak paramagnetic jumps responsible for the global polarization inversions.
The latter are paralleled by an increased jump rate as the jump history reveals (black clusters). Such intermittent fluctuations of the
jump rate are a well-known phenomenon of dynamical phase transitions in dissipative setups \cite{Ates2012,Lesanovsky2013}.
In the paramagnetic regime, the correlations vanish with $\kappa \to \infty$ due to frequent paramagnetic jumps, and the initial $m_z$-polarization is lost rapidly.
These observations support our claim of a non-equilibrium phase transition motivated by mean field calculations --- although the small system sizes
rendern any definite conclusion impossible.



Let us close this first part with a short résumé: We introduced a dissipative version of the transverse field Ising model
and showed that (1) we can probe the pure quantum phases of the Hamiltonian theory in the limiting regimes and (2) the mean field theory
predicts a non-equilibrium counterpart of the order-disorder phase transition.
Succeeding with this paradigmatic model rises the question whether more complex theories allow for an analogous
dissipative mimicry to probe their quantum phases and find interesting non-equilibrium phase transitions.
We answer in the affirmative, introducing the


\begin{figure*}[t]
  \includegraphics[width=1\linewidth]{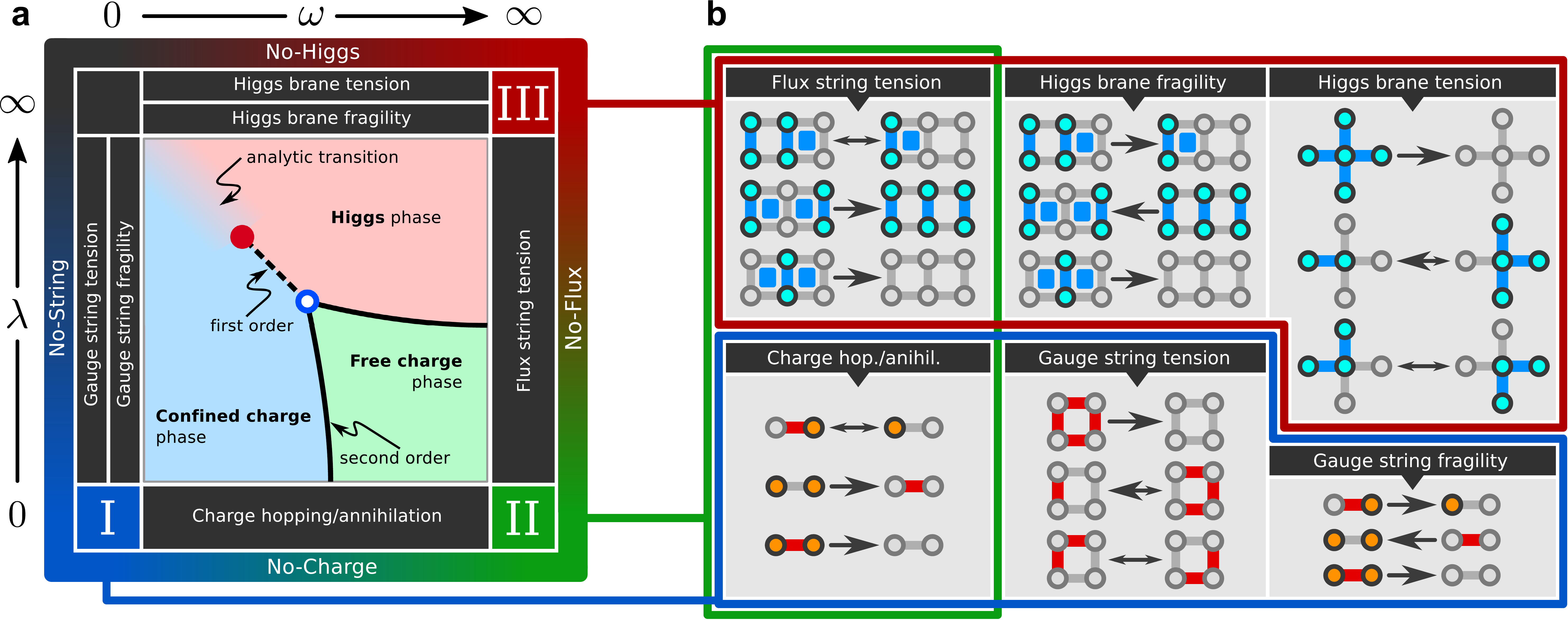}
  \caption{\label{fig:ghm_1} 
    Conceptual foundation of the dissipative $\mathbb{Z}_2$-Gauge-Higgs model. (A) illustrates qualitatively the well-known phase diagram of the Hamiltonian $\mathbb{Z}_2$-Gauge-Higgs theory
    in the $\omega$-$\lambda$-plane. There are three characteristic phases: The (I) confined charge, (II) free charge, and (III) Higgs phase. In order to drive the system dissipatively in a distinct phase,
    combinations of the baths adjacent to the labels (I), (II), and (III) are employed.
    (B) depicts the effects of the six types of jump operators (characterizing the baths) on elementary excitations in two spatial dimensions.
    Asymmetric arrows denote asymmetric quantum jump probabilities. The symbols read as follows: Yellow site $\Leftrightarrow$ $\sigma^x=-1$ (electric charge); Red edge $\Leftrightarrow$ $\tau^x=-1$ (gauge string); Blue site+edge $\Leftrightarrow$ $I_e=-1$ (Higgs excitation); Blue face $\Leftrightarrow$ $B_p=-1$ (magnetic flux).
    The formal definitions are given in Table~\ref{tab:LGT_setup}.
  }
\label{fig:LGT_setup}
\end{figure*}

\paragraph{Dissipative $\mathbb{Z}_2$-Gauge-Higgs model.}
Motivated by the possibility to explore quantum phases with driven dissipation, we present a dissipative implementation
of the famous $\mathbb{Z}_2$-Gauge-Higgs ($\mathbb{Z}_2$GH) model \cite{Wegner1971,Fradkin1978,Fradkin1979}.

Recently there has been intensified interest in the quantum simulation of gauge theories
\cite{Zohar2013,Stannigel2013,Marcos2014}, where the focus so far lies on the robust realization of the gauge constraints.
Here we do not focus on the latter but on the dynamics \textit{within the gauge invariant sector} itself.
%
%
To this end, consider a $D$-dimensional rectangular lattice with spin-$1/2$ representations attached to sites $s$ (the \textit{matter field}, denoted by $\sigma_s^k$)
and edges $e$ (the \textit{gauge field}, denoted by $\tau_e^k$). Here, $\sigma_s^k$ and $\tau_e^k$ ($k=x,y,z$) denote Pauli matrices. 
Then the Hamiltonian of the $\mathbb{Z}_2$GH model reads
\begin{equation}
 \label{eq:ghm_hamiltonian}
 H_{\mathbb{Z}_2\text{GH}}=-\sum_{s}\sigma_s^x-\lambda\sum_e I_e-\sum_{e}\tau_e^x-\omega\sum_{p}B_p
\end{equation}
where $s$, $e$ and $p$ denote sites, edges and faces of the (hyper-)cubic lattice, respectively; $\omega$ and $\lambda$ are non-negative real parameters.
The plaquette operators $B_p\equiv\prod_{e\in p}\tau_e^z$ 
describe a four-body interaction of gauge spins on the perimeter of face $p$ and
$I_e\equiv \sigma_{s_1}^z\tau_e^z\sigma_{s_2}^z$ (where $e=\{s_1,s_2\}$) realizes a gauged Ising interaction between adjacent matter spins.
Note that $H_{\mathbb{Z}_2\text{GH}}$ features the \textit{local} gauge symmetry $G_s\equiv \sigma_s^x\prod_{e:s\in e}\tau^{x}_e=\sigma_s^xA_s$, i.e. $\left[H,G_s\right]=0$ 
for all sites $s$. Here $A_s\equiv \prod_{e:s\in e}\tau^{x}_e$ denotes a $2D$-body interaction of gauge spins located on the edges adjacent to site $s$.


\begingroup
\renewcommand{\arraystretch}{1.6}
\begin{table}[b]
\begin{center}
\begin{ruledtabular}
\begin{tabular}{p{4cm} p{5cm}}
\textbf{Bath} & \textbf{Jump operator} \\ \hline
\noalign{\smallskip}
Gauge string tension & \small $F_p^{(1)}=\eta_1\,\,B_p\left(\mathds{1}-\tau_{e\in p}^x\right)$\\
Gauge string fragility & \small $F_e^{(2)}=\eta_2\,I_e\left(\mathds{1}-\tau_e^x\right)$\\ \hline
Higgs brane tension & \small $D_s^{(1)}=\eta_3\,\sigma_s^x\left(\mathds{1}-I_{e\in s}\right)$\\
Higgs brane fragility & \small $D_e^{(2)}=\eta_4\,\tau_e^x\left(\mathds{1}-I_e\right)$\\\hline
Charge hopping & \small $T_e=\eta_5\,I_e\left(\mathds{1}-\sigma_{s\in e}^x\right)$\\\hline
Flux string tension & \small $B_e=\eta_6\,\tau_e^x\left(\mathds{1}-B_{p\in e}\right)$\\
\end{tabular}
\end{ruledtabular}
\end{center}
\caption{Jump operators for the dissipative $\mathbb{Z}_2$-Gauge-Higgs model. Their action is described in the text. 
  Pictorial descriptions can be found in Fig.~\ref{fig:LGT_setup}. 
  $i$, $e$ and $p$ denote sites, edges and faces, respectively. The short-hand notation $e\in p$ denotes the normalized sum over all edges $e$ adjacent to face $p$. 
  The free parameters of the theory are labeled $\eta_i$ for $i=1,\dots,6$.
  The second column lists the jump operators of the gauge theory with non-trivial gauge condition 
  $\sigma_s^xA_s=\mathds{1}$.}
\label{tab:LGT_setup}
\end{table}
\endgroup

The expected quantum phase diagram in $2+1$ dimensions is sketched in Fig.~\ref{fig:LGT_setup}~(A) and features three distinct phases 
\cite{Fradkin1979,Tupitsyn2010}: The (I) confined charge, (II) free charge, and (III) Higgs phase, respectively.
To contrive a family of baths that explore these three phases and give rise to a non-equilibrium analogy of Fig.~\ref{fig:LGT_setup}~(A), it proves advantageous
to analyse the elementary excitations of $H_{\mathbb{Z}_2\text{GH}}$ in the three parameter regimes: We aim at jump operators that remove the elementary excitations
of each phase and thereby drive the system towards the latter. In addition, this approach leads inevitably to gauge invariant
jump operators $L$, i.e. $\com{L}{G_s}=0$ for all sites $s$.
For the sake of brevity, we label localised excitations (``quasiparticles'') by the corresponding operator in Hamiltonian~(\ref{eq:ghm_hamiltonian}) and its eigenvalue.
E.g. $\sigma_s^x=-1$ refers to a state $\Ket{\chi}$ such that $\sigma_s^x\Ket{\chi}=-\Ket{\chi}$ and we say that $\Ket{\chi}$ describes a system with an (electric) charge
at site $s$.

\begin{figure*}[t]
  \includegraphics[width=0.95\linewidth]{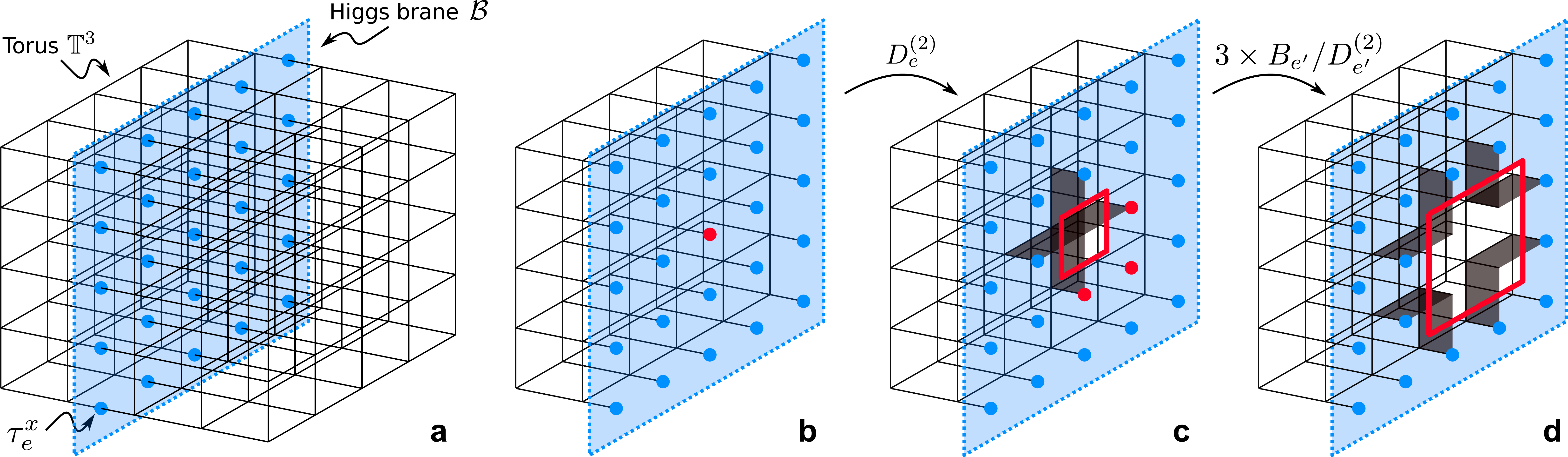}
  \caption{
    Action of the Higgs brane fragility $D_e^{(2)}$ and the Flux string tension $B_e$ in three spatial dimensions. 
    (A) Closed dual plane (Higgs brane) $\mathcal{B}$ on the three dimensional torus $\mathbb{T}^3$ which defines the topologically non-trivial brane operator
    $\prod_{e\in\mathcal{B}}\tau_e^x$ (blue points). Such excitations cannot be annihilated by 
    \textit{Higgs brane tension} $D_s^{(1)}$ and \textit{Flux string tension} $B_e$ since $\mathcal{B}$ wraps once around the torus and there are no flux strings present.
    (B-D) illustrates a cross-section of the lattice parallel to $\mathcal{B}$; we show the first few jumps to get rid of the excitations:
    (B) A \textit{Higgs brane fragility} jump $D_e^{(2)}$ acts on the red edge $e$. (C) The Higgs excitation $I_e=-1$ is no longer present. One has to pay for this with
    four magnetic fluxes $B_p=-1$ on the adjacent faces $p\in e$ (black plaquettes). The latter define a flux loop (red line).
    (D) Three applications of \textit{flux string tension} $B_{e^\prime}$ and/or \textit{Higgs brane fragility} $D_{e^\prime}^{(2)}$ on the edges $e^\prime$ (red points)
    enlarge the hole in the Higgs brane. This process allows the retraction and subsequent annihilation of the formerly closed Higgs brane around the torus.
  }
  \label{fig:LGT_fragility}
\end{figure*}

We start with the confined charge phase (I) for $\lambda,\omega\to 0$. The Hamiltonian reads 
$H_{\mathbb{Z}_2\text{GH}}=-\sum_{s}\sigma_s^x-\sum_{e}\tau_e^x$ and the elementary excitations are charges $\sigma_s^x=-1$
and gauge strings $\tau_s^x=-1$. The physically admissible, that is, \textit{gauge invariant} excitations are generated by $I_e$ and $B_p$, where $I_e$
creates a pair of charges on adjacent sites connected by a gauge string (usually called a \textit{meson}) and $B_p$ gives rise to a closed gauge string on the perimeter
of $p$. We conclude that physical states are characterized by (1) closed gauge strings and (2) open gauge strings with charges attached to their endpoints.
Such states obey a Gauss-like law, $G_s\Ket{\chi}=\Ket{\chi}$ for all $s$, which restricts the physical states to the gauge invariant subspace of the complete Hilbert space
characterized by $\sigma_s^xA_s=\mathds{1}$.
Note that the energy for separating two charges grows linearly with their distance since gauge strings are penalized by the Hamiltonian; thus the charges are \textit{confined}
whichs gives rise to the name \textit{confined charge phase}.

Let us now shift attention to the \textit{dissipative} analogue theory.
To get rid of an arbitrary configuration of charges (confined by gauge strings) and gauge loops, 
a \textit{gauge symmetric} dissipative process must (1) contract gauge strings, (2) annihilate pairs of charges, and (3) break gauge loops by creating mesons. 
The latter is only necessary for systems with non-trivial spatial topology, e.g. systems with periodic boundary conditions. We end up with the three baths 
\textit{Charge hopping/annihilation}, \textit{Gauge string tension}, and \textit{Gauge string fragility}, see Fig.~\ref{fig:LGT_setup}~(B) for a pictorial description
and Tab.~(\ref{tab:LGT_setup}) for the formal definition of the jump operators.

\begin{figure*}[t]
  \includegraphics[width=0.95\linewidth]{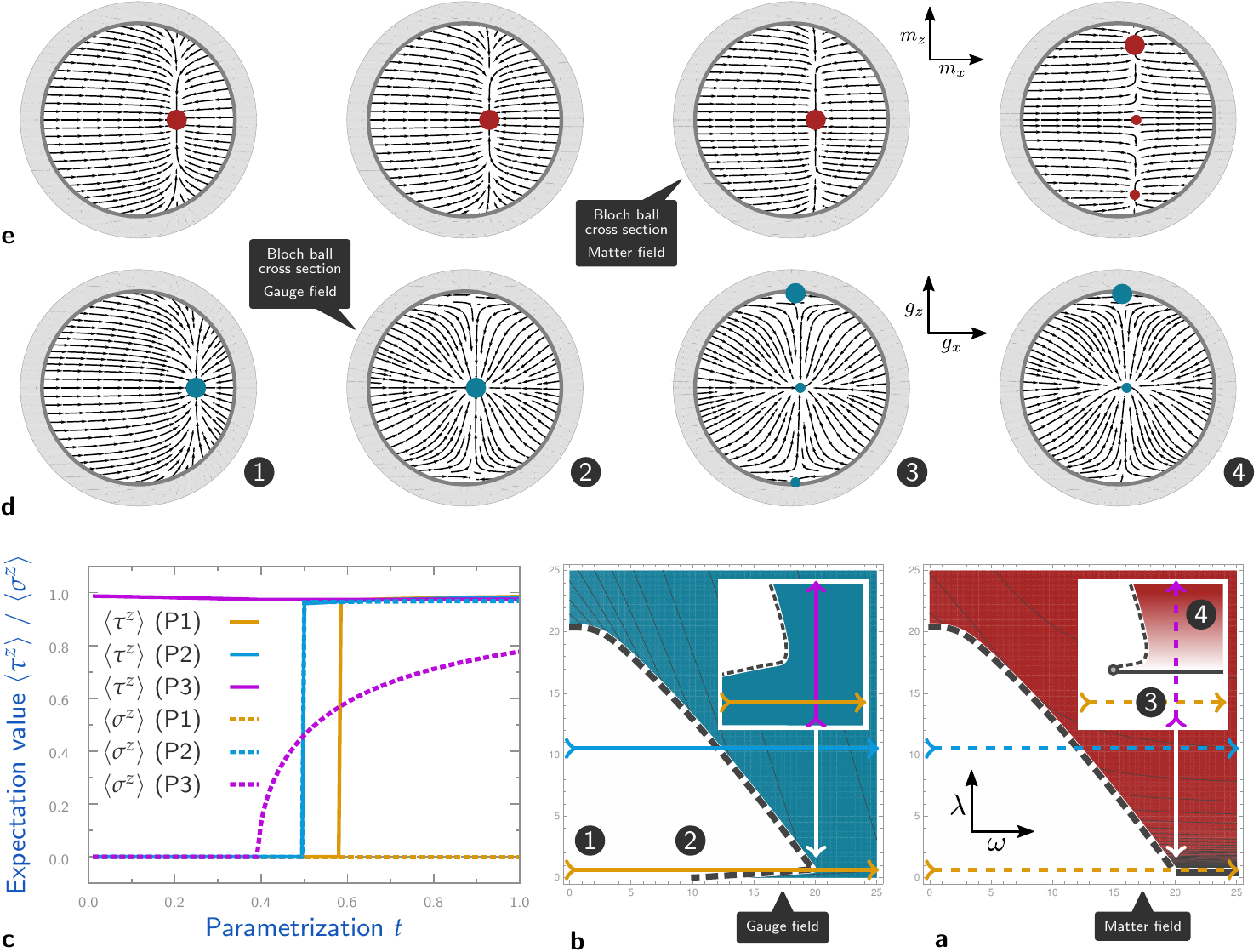}
  \caption{
    Mean field phase diagram for the dissipative $\mathbb{Z}_2$-Gauge-Higgs model with two separate mean fields. 
    In (A) we plot the maximal $z$-polarisation $\langle\sigma^z\rangle$
    of all \textit{stable physical steady states} for the matter field colour-coded in the $\omega$-$\lambda$-plane (light $\rightarrow$ $\langle\sigma^z\rangle=0$, dark $\rightarrow$ $\langle\sigma^z\rangle=1$).
    (B) shows the same for the gauge field, that is, $\langle\tau^z\rangle$.
    (C) depicts the quantitative results for $\langle\tau^z\rangle$ (solid) and $\langle\sigma^z\rangle$ (dashed) on the coloured paths in (B) and (A). 
    (D) and (E) illustrate cross sections of the Bloch ball ($g_x/m_x$-$g_z/m_z$-plane) for the gauge (D) and matter field (E) with the dynamical mean field flow $\vec F$ as flux lines and the stable physical fixed points marked by (small and large) discs.
    The corresponding parameters $(\omega,\lambda)$ for each vertical pair of cross sections are highlighted by numbers in the 2D plots (A) and (B).
    The shown flux lines for the Bloch vector of one mean field depend on the Bloch vector of the other since the mean field equations couple all six degrees of freedom. Each depicted gauge field flux corresponds to
    a fixed point Bloch vector for the corresponding matter field and vice versa (marked by large red and cyan discs).
    A discussion of the results is given in the text.
  }
  \label{fig:LGT_plots_A}
\end{figure*}

We proceed with the discussion of the remaining two phases.
The free charge phase (II) is characterized by $\lambda\to 0$ and $\omega\to\infty$ and the system is described by 
$H_{\mathbb{Z}_2\text{GH}}=-\sum_{s}\sigma_s^x-\omega\sum_{p}B_p$. Clearly, the matter and the gauge field decouple and the elementary excitations
are charges $\sigma_s^x=-1$ and magnetic fluxes $B_p=-1$ as excitations of the gauge string condensate. The latter appear as deconfined magnetic monopoles
in $D=2$ at the end of dual $\tau_e^x$-strings and as closed magnetic flux strings in $D=3$ on the perimeter of dual $\tau_e^x$-planes \footnote{The deconfinement of \textit{both}, charges and fluxes, in two spatial dimensions is directly related to the thermal instability of the toric code.}. 
Note that the charges are still created in pairs by $I_e$-chains; the connecting gauge strings however are no longer penalised, hence \textit{free} charge phase.
We conclude that the jump operators must provide mechanisms (1) to diffuse and annihilate charges and (2)
to do the same with magnetic mononpoles in $D=2$ and contract magnetic flux strings in $D=3$. This leads us to the already known \textit{Charge hopping/annihilation}
and the new \textit{Flux string tension} (which degenerates in $D=2$ to ``Monopole hopping/annihilation''), see Fig.~\ref{fig:LGT_setup}~(B) and Tab.~(\ref{tab:LGT_setup}).

Finally, the Higgs phase (III) is reached for $\lambda,\omega\to\infty$ and the Hamiltonian reads $H_{\mathbb{Z}_2\text{GH}}=-\lambda\sum_e I_e-\omega\sum_{p}B_p$. 
The elementary excitations are Higgs excitations $I_e=-1$ and flux strings $B_p=-1$. Pure Higgs excitations can be created by $\sigma_s^x$ and form dual loops in
$D=2$ and closed dual surfaces (``branes'') in $D=3$. Flux strings can be created by dual strings of $\tau_e^x$ or dual planes with boundary of $\tau_e^x$ in $D=3$.
That is, magnetic fluxes (as monopoles in $D=2$ or flux strings in $D=3$) mark the boundary of (dual) Higgs excitation manifolds, i.e. open strings in $D=2$ and open
branes in $D=3$.
Since in two dimensions the flux strings degenerate to magnetic monopoles, the physics becomes dual to the free charge phase (I)
via the identifications $\sigma_s^x\leftrightarrow B_p$ and $\tau_e^x\leftrightarrow I_e$. 
This duality should be preserved in our analogous dissipative setup.
Appropriate dissipative processes must (1) get rid of the flux strings/monopoles and (2) eliminate the Higgs excitations.
We handle the flux strings/monopoles by the already known \textit{Flux string tension} and introduce two new baths, the \textit{Higgs brane tension}
and the \textit{Higgs brane fragility}, to eliminate pure Higgs excitations. Since Higgs excitations can be created by both, $\sigma_s^x$ and $\tau_e^x$, in
the form of closed branes, the latter must be contracted and \textit{cut} in order to vanish on non-trivial topologies.
The cutting of Higgs branes is indeed necessary in three dimensions since topologically non-trivial, dual brane operators $\prod_{e\in\mathcal{B}}\tau_e^x$ ($\mathcal{B}$ is a dual plane that winds once around the torus $\mathbb{T}^3$) 
create excitation patterns that can only be annihilated by ``piercing holes'' in the Higgs brane to retract it about $\mathbb{T}^3$, see Fig.~(\ref{fig:LGT_fragility}).
The above mentioned duality in two dimensions becomes manifest in the duality relating \textit{Higgs brane fragility} and \textit{gauge string fragility}.
This becomes particularly clear in the ($D=2$) pictorial representations of Fig.~\ref{fig:LGT_setup}~(B).

At this point it seems advisable to stress the differences between the Hamiltonian theory and its dissipative counterpart. Ground states of the Hamiltonian theory
minimize the free energy, or, at zero temperature, the energy of the system. To reach, say, the quantum phase at $T=0$, the Hamiltonian system is coupled to
a thermal bath whose temperature is gradually reduced towards zero. The cooling of the system is driven by \textit{thermal fluctuations} which are conditioned according
their Boltzmann weight with respect to the system Hamiltonian. It is important to stress that whether a certain transformation occurs 
(e.g. the breaking of a gauge loop into an open gauge string with charges terminating the strings) depends solely on its energetic effect with respect to the Hamiltonian.
In contrast, there is no such thing as energy in the dissipative non-equilibrium setup. Consequently, the options for microscopic fluctuations are much more constrained,
namely by the possible actions of the jump operators. \textit{Dissipative fluctuations are transformation-selective whereas thermal fluctuations are energy-selective.}
Consider once again the breaking of gauge loops: In a (thermal) Hamiltonian theory they will just break whenever it is energetically favourable. In our purely dissipative
setup they can only break if we allow them to do so, that is, if we provide an appropriately designed bath with jump operators that break strings
(in our case this bath is termed \textit{gauge string fragility} and controlled by the parameter $\eta_2$, see Tab.~(\ref{tab:LGT_setup})).
To put it in a nutshell, the translation of Hamiltonian ``blue print'' theories into a purely dissipative non-equilibrium framework allows for much more
fine-tuning on the microscopic level.

The relative bath strengths $\eta_i$, $i=1,\dots,6$, (see Tab.~(\ref{tab:LGT_setup})) are free parameters of our theory and allow for the mentioned 
fine tuning of the microscopic mechanisms. For instance, there is no a priori statement about the importance of ``gauge string breaking'' as compared to 
``gauge string tension'' and the influence of such ratios on the phase diagram is highly non-trivial. 
However, in the following we set $\eta_{1,2}=1=\eta_5$, $\eta_{3,4}=\sqrt{\lambda}$ and $\eta_6=\sqrt{\omega}$ since this seems a natural choice to mimic the 
original theory~(\ref{eq:ghm_hamiltonian}).

\paragraph{Mean field analysis.}

To put the theory into operation and catch a glimpse at its qualitative phase diagram, we once again utilize a mean field approach.
Mean field approximations for theories with (unphysical) gauge degrees of freedom
are well known to yield not only quantitatively poor but also qualitatively wrong results~\cite{Drouffe1983,Dagotto1984,Alvarez1986}.
However, we can test the ability of our \textit{dissipative} $\mathbb{Z}_2$GH model to realize the different quantum phases of 
the \textit{Hamiltonian} $\mathbb{Z}_2$GH theory by comparing the predictions of both models within mean field theory, where the features and shortcomings
for the \textit{Hamiltonian} $\mathbb{Z}_2$GH model are well established \cite{Drouffe1983,Dagotto1984}.


We followed two different mean field approaches, the combination of which is known to capture 
all essential features of the quantum phase diagram for the Hamiltonian theory. The results for one of these approaches
are shown in Fig.~\ref{fig:LGT_plots_A} and we find that they correspond qualitatively to the results of the Hamiltonian counterpart.
An alternative approach in unitary gauge is discussed in the methods section.

Here we present the simplest approach to obtain an effective mean field description of the theory by introducing \textit{two} independent mean field degrees
of freedom. That is, we make the ansatz
\begin{equation}
  \rho=\bigotimes_{e\in\mathbb{E}}\rho^{\text{g}}_e\otimes\bigotimes_{s\in\mathbb{S}}\rho^{\text{m}}_s
\end{equation}
for the density matrix, where $\rho^{\text{g}}_e=(\mathds{1}_e+\vec g\vec\tau_e)/2$ 
describes the single-site gauge field with Bloch vector $\vec g=(g_x,g_y,g_z)$ and
$\rho^{\text{m}}_s=(\mathds{1}_s+\vec m\vec\sigma_s)/2$ 
analogously the matter field with Bloch vector $\vec m=(m_x,m_y,m_z)$.
Self-consistency once again demands $g_k=\langle \tau_e^k\rangle$ and $m_k=\langle \sigma_s^k\rangle$ for $k=x,y,z$;
assuming a homogeneous system allows us to omit the site and edges indices.
An analogous treatment as in the case of the dissipative TIM yields a non-linear dynamical system with the $6$-dimensional flow 
$\vec F(\vec g,\vec m)=(\vec F^{\text{g}},\vec F^{\text{m}})$, namely
\begin{equation}
  \partial_t\vec g=\vec F^{\text{g}}(\vec g,\vec m)\quad\text{and}\quad
\partial_t\vec m=\vec F^{\text{m}}(\vec g,\vec m)\,.
\end{equation}
Stationary states (NESS) can be determined by solving the non-linear system of equations 
$\vec F^{\text{g}}(\hat{\vec g},\hat{\vec m})=0=\vec F^{\text{m}}(\hat{\vec g},\hat{\vec m})$
and their stability can be infered from the spectrum of $\mathrm{D}\vec F(\hat{\vec g},\hat{\vec m})$.

The results are shown in Fig.~\ref{fig:LGT_plots_A}.
In (A) and (B) we illustrate the expectation values $m_z=\langle\sigma^z\rangle$ and $g_z=\langle\tau^z\rangle$ for
the matter and the gauge field, respectively; (C) shows these quantities on the three highlighted paths.
In the case of multiple stable solutions, we choose the one which maximises first $g_z$, and then $m_z$.
For the Hamiltonian mean field approach such a selection can be justified by comparing the free energies of all possible solutions.
Lacking an extremum principle in the non-equilibrium setting, it remains an open question which solutions are truly stable and which, in contrast, give
rise to metastable states (or do not exist at all).

Nevertheless we find three distinct phases, characterized by the existence of solutions with $m_z=0=g_z$ (1 and 2), $m_z=0\neq g_z$ (3), and $m_z\neq 0\neq g_z$ (4).
They can be identified with the confined charge, free charge, and Higgs phase, respectively. There are two types of phase transitions present, see (C).
The confined charge phase is separated from the other two phases by a first order transition which is indicated by a jump $g_z=0\to g_z>0$,
the transition between free charge and Higgs phase is of second order and indicated by a continuous transition $m_z=0\nearrow m_z>0$.

We have to lower our sights regarding the graphical representation of the $6$-dimensional mean field flow $\vec F(\vec g,\vec m)=(\vec F^{\text{g}},\vec F^{\text{m}})$
in (D) and (E). Here we show (the projection of) $\vec F^{\text{g}}(\vec g,\hat{\vec m})$ in (D) and $\vec F^{\text{m}}(\hat{\vec g},\vec m)$ in (E) for the fixed
points $\hat{\vec g}$ and $\hat{\vec m}$ marked by bold disks in the corresponding cross section. Other stable fixed points are labeld by small disks of the same colour.
In the confined charge phase there is a unique stable fixed point, see (1) and (2). In the free charge phase two additional stable fixed points emerge close to the 
$g_z=\pm 1$ poles which are responsible for the first order phase transition. All three stable solutions correspond to a vanishing matter field $m_z=0$.
In the Higgs phase the solution close to the $g_z=-1$ pole vanishes and only the ones close to $g_z=0$ and $g_z=1$ remain. There are three solutions, namely
$g_z=0=m_z$, $g_z>0< m_z$, and $g_z>0>m_z$. That the solutions of the gauge field are \textit{not} symmetric about the $g_x$-axis (horizontal axis in the cross sections)
whereas the matter field solutions feature this symmetry about the $m_x$-axis is related to the fact, that the theory features the global symmetry 
$\prod_s\sigma_s^x=\prod_s G_s$ but \textit{not} an analogous symmetry $\prod_e\tau_e^x$ for the gauge field.

An obvious drawback of this mean field approach is that the gauge degrees of freedom are \textit{not} fixed and erroneously treated as physical degrees of freedom. 
This leads to the well-known artifact that the analytical path connecting confined charge and
Higgs phase is lost. However, the theory predicts all \textit{three} phases correctly.

To properly exclude \textit{unphysical} degrees of freedom, it proves advantageous to localise the latter on distinguished \textit{mathematical} degrees of freedom.
This can be achieved in unitary gauge where the physical subspace $\mathcal{H}_{\mathbb{Z}_2GH}=\{\Ket{\Psi}\,|\,G_s=\mathds{1}\}$ is unitarily rotated into
the new subspace $\tilde{\mathcal{H}}_{\mathbb{Z}_2GH}=\{\Ket{\Psi}\,|\,\sigma_s^x=\mathds{1}\}=T\mathcal{H}_{\mathbb{Z}_2GH}$ via $T$.
Then one finds a first order phase transition separating confined charge and free charge \& Higgs phase --- the latter two being no longer distinct.
In contrast to our approach above, the first order line \textit{terminates} at a critical point $(\omega_c,\lambda_c)$ and the analytical transition
of Fig.~\ref{fig:LGT_setup}~(A) is recovered within mean field theory.
These results once again parallel the already known mean field phase diagram of the Hamiltonian theory in unitary gauge \cite{Drouffe1983,Dagotto1984}.
For a detailed discussion, the reader is referred to the methods section.


\paragraph{Discussion.}

In this manuscript we introduced the mimicry of well-known (quantum) phase transitions
by Markovian non-equilibrium systems. We illustrated the construction of competing
baths for a simple paradigmatic system --- the transverse field Ising model ---
and the considerably more complex $\mathbb{Z}_2$ lattice gauge theory with coupled matter field.
For this purpose we employed the Hamiltonian versions of the theories as ``blue prints''
to come up with appropriate jump operators that drive the dissipative system
towards the pure quantum phases of the Hamiltonian theory.
We pointed out that the non-equilibrium framework can be seen as a
``construction kit'' for phase transitions that features more control
over the microscopic behaviour than any Hamiltonian theory by probing the much richer non-thermal manifold of states. 
We believe that such purely dissipative quantum simulations can serve as a new, generic and inherently robust tool for the exploration of 
otherwise inaccessible phase diagrams of complex quantum systems.



\appendix

\begin{figure*}[t]
  \includegraphics[width=0.95\linewidth]{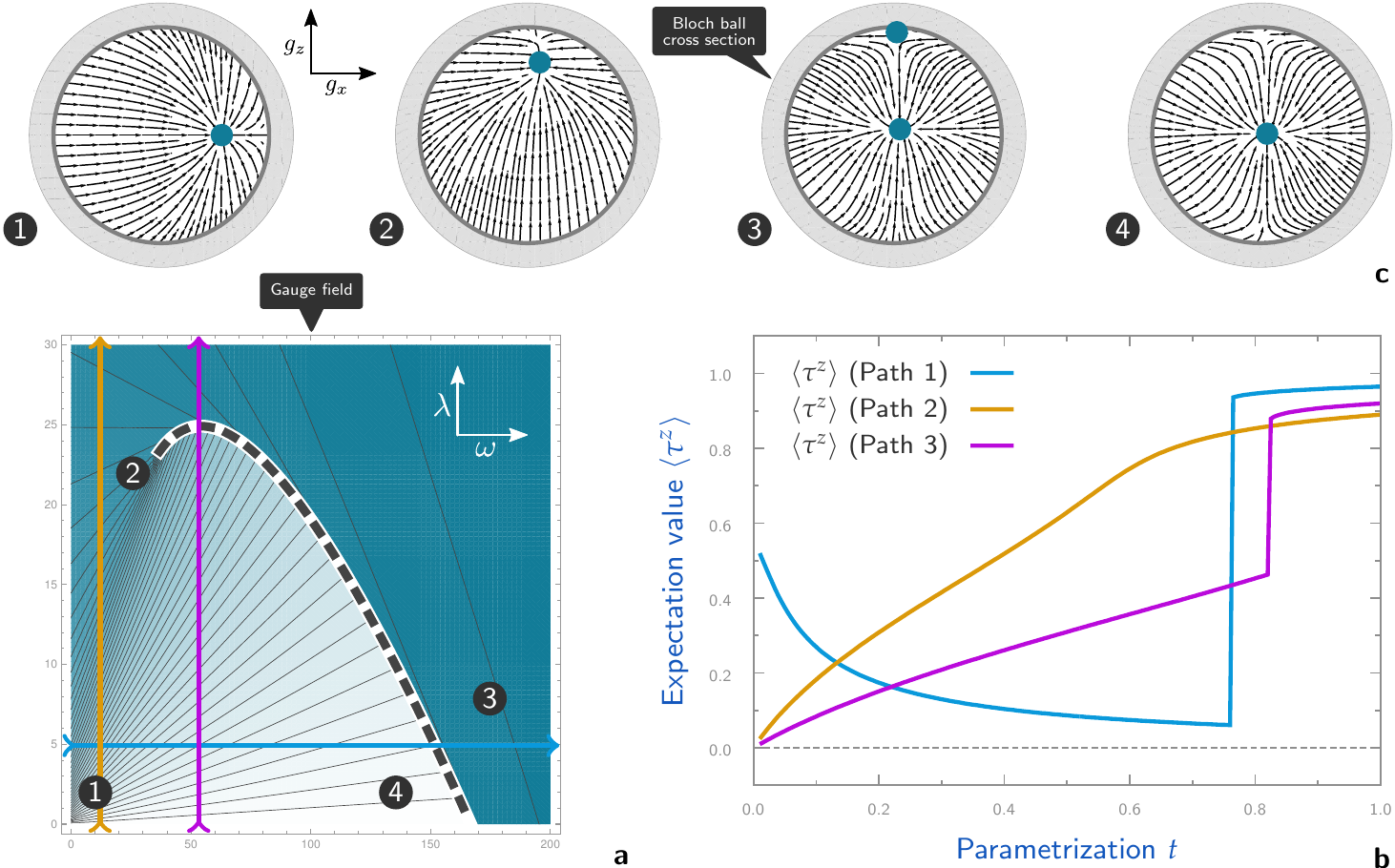}
  \caption{
    Mean field phase diagram for the dissipative $\mathbb{Z}_2$-Gauge-Higgs model in unitary gauge. 
    In (A) we plot the maximal $z$-polarisation $\langle\tau^z\rangle$ of all \textit{stable physical steady states}
    colour-coded in the $\omega$-$\lambda$-plane (light $\rightarrow$ $\langle\tau^z\rangle=0$, dark $\rightarrow$ $\langle\tau^z\rangle=1$).
    In (B) we show the quantitative results for $\langle\tau^z\rangle$ on the coloured paths in (A). (C) illustrates four characteristic cross sections
    of the Bloch ball ($m_x$-$m_z$-plane) with the dynamical mean field flow $\vec F$ as flux lines and the stable physical fixed points marked by cyan discs.
    The corresponding parameters $(\omega,\lambda)$ for each cross section are highlighted by numbers in the 2D plot (A).
    A discussion of the results is given in the text.
  }
  \label{fig:LGT_plots_B}
\end{figure*}

\begingroup
\renewcommand{\arraystretch}{1.6}
\begin{table*}[t]
  \begin{center}
    \begin{ruledtabular}
      \begin{tabular}{p{4cm} p{5cm} p{5cm}}
        \textbf{Bath} & \textbf{Gauge condition $\sigma_s^xA_s=\mathds{1}$} & \textbf{Gauge condition $\sigma_s^x=\mathds{1}$} \\ \hline
        \noalign{\smallskip}
        Gauge string tension & \small $F_p^{(1)}=\eta_1\,\,B_p\left(\mathds{1}-\tau_{e\in p}^x\right)$ & \small $\tilde F_p^{(1)}=\eta_1\,B_p\left(\mathds{1}-\tau_{e\in p}^x\right)$\\
        Gauge string fragility & \small $F_e^{(2)}=\eta_2\,I_e\left(\mathds{1}-\tau_e^x\right)$ & \small $\tilde F_e^{(2)}=\eta_2\,\tau_e^z\left(\mathds{1}-\tau_e^x\right)$\\ \hline
        Higgs brane tension & \small $D_s^{(1)}=\eta_3\,\sigma_s^x\left(\mathds{1}-I_{e\in s}\right)$ & \small $\tilde D_s^{(1)}=\eta_3\,A_s\left(\mathds{1}-\tau_{e\in s}^z\right)$\\
        Higgs brane fragility & \small $D_e^{(2)}=\eta_4\,\tau_e^x\left(\mathds{1}-I_e\right)$ & \small $\tilde D_e^{(2)}=\eta_4\,\tau_e^x\left(\mathds{1}-\tau_e^z\right)$\\ \hline
        Charge hopping & \small $T_e=\eta_5\,I_e\left(\mathds{1}-\sigma_{s\in e}^x\right)$ & \small $\tilde T_e=\eta_5\,\tau_e^z\left(\mathds{1}-A_{s\in e}\right)$\\ \hline
        Flux string tension & \small $B_e=\eta_6\,\tau_e^x\left(\mathds{1}-B_{p\in e}\right)$ & \small $\tilde B_e=\eta_6\,\tau_e^x\left(\mathds{1}-B_{p\in e}\right)$\\
      \end{tabular}
    \end{ruledtabular}
  \end{center}
  \caption{Jump operators for the dissipative $\mathbb{Z}_2$-Gauge-Higgs model (comparison). Their action is described in the main text. 
    Pictorial descriptions can be found in Fig.~\ref{fig:LGT_setup} of the main text. 
    $i$, $e$ and $p$ denote sites, edges and faces, respectively. The short-hand notation $e\in p$ denotes the normalized sum over all edges $e$ adjacent to face $p$. 
    The free parameters of the theory are labeled $\eta_i$ for $i=1,\dots,6$.
    The centered column lists the jump operators of the gauge theory with non-trivial gauge condition 
    $\sigma_s^xA_s=\mathds{1}$ (Gauss law). A unitary transformation maps the theory to a new subspace which is defined by the trivial gauge condition $\sigma_s^x=\mathds{1}$.
    The jump operators in this subspace (unitary gauge) are listed in the right-hand column. The transformation is described in the text.}
  \label{tab:LGT_setup_large}
\end{table*}
\endgroup

\section{Methods}

\subsection{Mean field theory for Lindblad master equations}
\label{app:mf_theory}

\paragraph{Mean field jump operators.}
Let the system's states be described by the $N$-spin 
Hilbert space $\mathcal{H}_N=\bigotimes_{i=1}^N\mathbb{C}^2_i$.
For mean field theory we choose the ansatz 
$\rho=\bigotimes_l\rho_l$ where $\rho_l$ is the density matrix of a single spin degree of freedom.
Here we consider the generic case, that is, we allow for $1\leq M \leq N$ independent spins in the
mean field description. E.g. for $M=1$ we end up with a completely 
homogeneous system; $M=N$ describes a system of $N$ distinguished spins which are incoherently
coupled to their neighbours via their expectation values.
Usually one will choose $\mathcal{O}(1)$ mean fields to assign a distinct mean field degree of freedom 
to all distinguished fields in the exact theory \footnote{For instance, 
  consider the  $\mathbb{Z}_2$-Gauge-Higgs model. Here one naturally introduces two mean fields 
  for the gauge and the matter field, respectively.}.

Given $M$ mean fields, the density matrix reads $\rho^\text{mf}=\bigotimes_{\alpha=1}^M\tilde\rho_\alpha$ 
where $\tilde\rho_\alpha$ describes the (homogeneous) $\alpha$-th mean field. 
The effective jump operators are obtained by tracing out selectively all degrees of freedom but one, meaning
\begin{equation}
  \label{eq:mf_trace}
  \partial_t\tilde\rho_\alpha=\partial_t\Tr{\rho}{\neq m}=\Tr{\mathcal{L}[\rho]}{\neq m}
\end{equation}
where $1\leq m\leq N$ is a physical spin which represents the field of type $\alpha$.
The dynamics of the mean field spins $\{\tilde\rho_\alpha\}$ is described by 
effective Lindblad equations
\begin{equation}
  \label{eq:mf_time_old}
  \partial_t\tilde\rho_\alpha=\sum_i\sum_{\mu_i}\left[l^{\alpha}_{i,\mu_i}\tilde\rho_\alpha {l^{\alpha}_{i,\mu_i}}^\dag
    -\frac{1}{2}\acom{{l^{\alpha}_{i,\mu_i}}^\dag l^{\alpha}_{i,\mu_i}}{\tilde\rho_\alpha}\right]
\end{equation}
where one has to keep in mind that these equations are \textit{non-linear} 
due to the mean fields included in the effective jump operators:
\begin{equation}
  L_i\,\stackrel{\alpha}{\longrightarrow}\, \left\{l^{\alpha}_{i,\mu_i}\right\}_{\mu_i}=\left\{l^{\alpha}_{i,\mu_i}\left(\left\{m^k_{\beta}\right\}\right)\right\}_{\mu_i}
\end{equation}
Here $m^k_{\beta}\equiv \langle \sigma^k_\beta \rangle=\tr{\sigma^k_\beta\tilde\rho_\beta}$ denotes the 
$k$-th component of the $\beta$-th mean field ($k=x,y,z$).
Furthermore notice that for each exact jump operator $L_i$ there may be several 
effective jump operators $l^{\alpha}_{i,\mu_i}$ with $\mu_i=1,2,3,\dots$ for each mean field $\alpha$.

For the sake of simplicity we employ a resummation and redefinition of the effective jump operators to get rid of duplicates
(which usually occur due to structural symmetries of the lattice). So rewrite Eq.~(\ref{eq:mf_time_old}) as
\begin{equation}
  \label{eq:mf_time}
  \partial_t\tilde\rho_\alpha=\sum_{\mu}\left[l^{\alpha}_{\mu}\tilde\rho_\alpha {l^{\alpha}_{\mu}}^\dag
    -\frac{1}{2}\acom{{l^{\alpha}_{\mu}}^\dag l^{\alpha}_{\mu}}{\tilde\rho_\alpha}\right]
\end{equation}
for the effective Markovian dynamics. The number of effective jump operators $\left\{l_\mu^\alpha\right\}$ is bounded and does not
depend on the system size $N$ (otherwise a mean field approximation would hardly be legitimate).
This is our starting point for the following analysis of non-equilibrium dynamics and steady states.

\paragraph{Dynamics.}

The generic form for single-spin mean field jump operators is
\begin{equation}
  l_\mu^\alpha = \sum_{\lambda=0}^3 l^{\alpha}_{\mu,\lambda} \sigma_{\alpha}^\lambda
  \quad\text{where}\quad
  l^{\alpha}_{\mu,\lambda}=l^{\alpha}_{\mu,\lambda}\left(\left\{m^k_{\beta}\right\}\right)\,.
\end{equation}
Henceforth we use Einstein's convention for Latin indices but not for Greek indices.
In the most generic case, jump operators are not traceless, i.e. $l^{\alpha}_{\mu,0}\neq 0$ 
(recall that $\sigma_{\alpha}^0 =\mathds{1}_\alpha$). 
However, in the models considered here these components vanish altogether 
and thus we assume $l^{\alpha}_{\mu,0}= 0$ henceforth.
To make this clear, we switch to Latin indices $i,j,k,\dots$ which run over $1,2,3$ 
(whereas Greek indices run over $0,1,2,3$ except for $\mu$ which indicates the different jump operators).

Let us introduce the three-index function
\begin{equation}
  \label{eq:mf_L}
  L_{i,j}^\alpha \equiv
  \sum_\mu\overline{l^{\alpha}_{\mu,i}} l^{\alpha}_{\mu,j}
  =R_{i,j}^\alpha + \imath I_{i,j}^\alpha
\end{equation}
with real part $R_{i,j}^\alpha=\Re L_{i,j}^\alpha$
and imaginary part $I_{i,j}^\alpha=\Im L_{i,j}^\alpha$.
Since $\vec L^\alpha=(L_{i,j}^\alpha)$ is a Hermitian matrix for all $\alpha$, we find
$R_{i,j}^\alpha=R_{j,i}^\alpha$ and $I_{i,j}^\alpha=-I_{j,i}^\alpha$ and thus
$R^\alpha\equiv L_{i,i}^\alpha=R_{i,i}^\alpha$. One may call $L^\alpha$ \textit{system matrices}
as they encode the complete mean field theory of the system.

Due to the product structure of $\rho^\text{mf}=\bigotimes_{\alpha=1}^M\tilde\rho_\alpha$ we can
parametrise each mean field density matrix as $\tilde\rho_\alpha=1/2\,
\left(\mathds{1}_\alpha+ a^k_\alpha\sigma_\alpha^k\right)$. Clearly, self-consistency requires
\begin{equation}
  m_\alpha^k=\tr{\sigma_\alpha^k\tilde\rho_\alpha}=a^k_\alpha
\end{equation}
so we can just substitute $a_\alpha^k$ by the expectation value 
$m_\alpha^k$, $\tilde\rho_\alpha=1/2\,\left(\mathds{1}_\alpha+ m^k_\alpha\sigma_\alpha^k\right)$.

With these definitions in mind it is straightforward to show that the mean field dynamics~(\ref{eq:mf_time}) 
is described by the set of generally non-linear differential equations
\begin{equation}
  \label{eq:mf_equation_unitary_dynamics}
  \partial_t m_\alpha^n
  =2\epsilon^{ijn} I^\alpha_{i,j}+2\left(R^\alpha_{n,i}-R^\alpha\delta_{ni}\right)m^i_\alpha
\end{equation}
where $\delta_{ni}$ denotes the Kronecker delta and $\epsilon^{ijn}$ the Levi-Civita symbol. 
If we consider all $m_\alpha^n$ ($\alpha=1,\dots,M$ and $n=1,2,3$) as independent real coordinates 
in $\mathbb{R}^{3M}$, it is convenient to define the vector field
\begin{equation}
  \label{eq:mf_equation_unitary_dynamics_field}
  \left[F\left(\{m_\beta^i\}\right)\right]_{(\alpha,n)}\equiv
  2\epsilon^{ijn}I^\alpha_{i,j}+2\left(R^\alpha_{n,i}-R^\alpha\delta_{ni}\right)m^i_\alpha
\end{equation}
which is the flow that determines the time evolution via the dynamical system
\begin{equation}
  \label{eq:mf_equation_unitary_dynamics_2}
  \partial_t \vec M=\vec F\quad\text{with}\quad
  \vec M\equiv \left(m_\alpha^n\right)_{(\alpha,n)}\,.
\end{equation}
For example, in Fig.~\ref{fig:TIM_plot}~(B) of the main text we illustrate the flow $\vec F$ 
for the dissipative transverse field Ising model in the Bloch ball ($M=1$).

\paragraph{Steady states.}

The mean field steady states are given by the solutions of $0\stackrel{!}{=}\partial_t \vec M=\vec F$.
Then Eq.~(\ref{eq:mf_equation_unitary_dynamics}) yields the system of generally
non-linear equations
\begin{equation}
  \label{eq:mf_equation}
  R^\alpha m^n_\alpha=\epsilon^{ijn}I^\alpha_{i,j}+m^i_\alpha R^\alpha_{n,i}
\end{equation}
for $n=1,2,3$ and $1\leq\alpha\leq M$.
Its solutions $(\hat m_\alpha^n)$ determine the steady states via 
$\tilde\rho_\alpha^\text{NESS}=1/2\,\left(\mathds{1}_\alpha+ \hat m^k_\alpha\sigma_\alpha^k\right)$. 
The stability of these solutions can be inferred from the spectrum $\sigma [\mathrm{D}\vec F]$ of the 
derivative (Jacobian matrix $J_{\vec F}$)
\begin{equation}
  \mathrm{D}\vec F=J_{\vec F}\equiv 
  \left[\frac{\partial F_{(\alpha,n)}}{\partial m^k_\beta}\right]_{(\alpha,n),(\beta,k)}
\end{equation}
at the fixed points $(\hat m_\alpha^n)$. A solution with $\max \sigma [\mathrm{D}\vec F(\hat m_\alpha^n)]<0$ 
is \textit{stable} and the corresponding state $\tilde\rho_\alpha^\text{NESS}$ is considered a physically
relevant mean field steady state. On the contrary, solutions with
$\max \sigma [\mathrm{D}\vec F(\hat m_\alpha^n)]>0$ are not of physical relevance as their fixed
points are unstable at least in one direction of the parameter space $\mathbb{R}^{3M}$.

\paragraph{Application to the TIM.}

Here we consider exemplarily the paradigmatic dissipative transverse field Ising model. 
Its competing jump operators are defined in~(\ref{eq:tim_jops}) of the main text.
If we assume a homogeneous system with a single mean field degree of freedom 
$m_k\equiv m_\alpha^k=\langle\sigma_i^k\rangle$ for all $1\leq i\leq N$, 
Eq.~(\ref{eq:mf_trace}) yields the ferromagnetic mean field jump operators (here $p_1\equiv f_0$, see main text)
\begin{subequations}
  \begin{eqnarray}
    f_1=\sigma^x\left[\mathds{1}-m_z\sigma^z\right]\,&\Rightarrow&\,\vec l_{f_1}=[1,im_z,0]\\
    f_2=q^{-1/2}\,\overline{m}_z\,\sigma^y\,&\Rightarrow&\,\vec l_{f_2}=q^{-1/2}\,[0,\overline{m}_z,0]\\
    f_3=q^{-1/2}\,\sigma^z\,&\Rightarrow&\,\vec l_{f_3}=q^{-1/2}\,[0,0,1] \\
    p_1=\sqrt{\kappa}\,\sigma^z\left[\mathds{1}-\sigma^x\right]\,&\Rightarrow&\,\vec l_{p_1}=\kappa^{-1/2}/2\,[0,-i,1]
  \end{eqnarray}
\end{subequations}
with the coordinate representations $l_{\mu,i}$ ($\mu=f_1,f_2,f_3,p_1$ and $i=1,2,3$).
For the sake of brevity we introduced the coordination number $q\equiv 2D$ 
and $\overline{m}_z\equiv\sqrt{1-m_z^2}$.

Please note that $f_1$ and $p_1$ remain finite in the high-dimensional limit $D\to\infty$ 
whereas the $y$- and $z$-dephasing $f_2$ and $f_3$ become irrelevant for high-dimensional systems 
and affects the results only quantitatively.

We can now evoke Eq.~(\ref{eq:mf_L}) and~(\ref{eq:mf_equation_unitary_dynamics_field}) 
to derive the mean field flow in the Bloch ball
\begin{equation}
  \label{eq:tim_mf_flux_dissipative}
  \vec F(\vec m) =
  \begin{bmatrix}
    -m_x\,\left[2 \left((1-\frac{1}{q}) m_z^2+\frac{2}{q}\right)+\kappa\right]+\kappa \\
    -m_y \left[2 (1+\frac{1}{q}) 1+\frac{\kappa}{2}\right]\\
    2 (1-\frac{1}{q}) m_z \left(1-m_z^2\right) 1-m_z\frac{\kappa}{2}
  \end{bmatrix}
\end{equation}
with the triangular Jacobian matrix
{\small
  \begin{widetext}
    \begin{equation}
      \label{eq:tim_mf_dissipative_jacobian}
      \mathrm{D}\vec F(\vec m)=
      \left[
        \begin{array}{ccc}
          -2 \left[(1-\frac{1}{q}) m_z^2+\frac{2}{q}\right] 1-\kappa & 0 & - 4m_x m_z1 (1-\frac{1}{q}) \\
          0 & -2 (1+\frac{1}{q}) 1-\frac{\kappa}{2} & 0 \\
          0 & 0 & -2 (1-\frac{1}{q}) \left(3 m_z^2-1\right) 1-\frac{\kappa}{2}
        \end{array}
      \right]\,,
    \end{equation}
  \end{widetext}
}
the spectrum of which can be read off.

Computation of the fixed points $\hat{\vec m}$ via Eq.~(\ref{eq:mf_equation}) ---
or equivalently $\vec F=0$ --- yields the three solutions
\begin{subequations}
  \label{eq:tim_mf_dissipative_steady_states}
  \begin{eqnarray}
    \hat{\vec m}_\text{P}&=&\begin{bmatrix}\frac{\kappa q}{\kappa q+4} & 0 & 0\end{bmatrix}^T\\
    \hat{\vec m}_\text{F1}&=&\begin{bmatrix}\frac{2 \kappa  q}{(\kappa +4) q+4} & 0 & -\frac{1}{2} \sqrt{4-\frac{\kappa  q}{q-1}}\end{bmatrix}^T \\
    \hat{\vec m}_\text{F2}&=&\begin{bmatrix}\frac{2 \kappa  q}{(\kappa +4) q+4} & 0 & +\frac{1}{2} \sqrt{4-\frac{\kappa  q}{q-1}}\end{bmatrix}^T
  \end{eqnarray}
\end{subequations}
which can be classified as paramagnetic ($\text{P}$, $\hat m_z=0$) and 
ferromagnetic ($\text{F}_1$ and $\text{F}_1$, $\hat m_z\neq 0$) solutions.

Clearly, the ferromagnetic solutions $\text{F}_1$ and $\text{F}_2$ become real valued (and thereby valid
Bloch vectors) iff
\begin{equation}
  4-\frac{\kappa  q}{q-1}\geq 0\quad\Leftrightarrow\quad
  \kappa\leq\kappa_c\equiv 4\left(1-\frac{1}{q}\right)
\end{equation}
where $\kappa_c$ is the \textit{critical coupling}.
We want to stress that $\lim_{D\to\infty}\kappa_c=4>0$ --- 
that is, the mean field phase transition is stable in the high-dimensional limit.

At this point it remains to check which of the three solutions for $\kappa<\kappa_c$ are the physical ones.
To this end we have to plug the fixed points in the three eigenvalues of 
Eq.~(\ref{eq:tim_mf_dissipative_jacobian}). This yields for the paramagnetic solution
\begin{equation}
  \lambda_1^\text{P}
  =2-\frac{1}{2}\left(\kappa+\frac{4}{q}\right)\lessgtr 0\,,\quad\lambda_2^\text{P}<0\,,\quad\lambda_3^\text{P}<0\,.
\end{equation}
We see that $\hat{\vec m}_\text{P}$ becomes unstable for 
$\kappa<\kappa_c$ since then $\lambda_1^P>0$. The same procedure for the ferromagnetic solutions yields
\begin{equation}
  \lambda_1^\text{F}
  =\kappa -4\left(1-\frac{1}{q}\right)\lessgtr 0\,,\quad\lambda_2^\text{F}<0\,,\quad\lambda_3^\text{F}<0
\end{equation}
which leads us to the conclusion that they become stable the moment they become real-valued, 
namely for $\kappa<\kappa_c$ when $\lambda_1^F$ becomes negative.

These discussions establish the phase diagram in Fig.~\ref{fig:TIM_plot}~(A) as well as the
qualitative structure of the mean field flow in Fig.~\ref{fig:TIM_plot}~(B) of the main text.

\subsection{Mean field theory for the $\mathbb{Z}_2$-Gauge-Higgs model in unitary gauge}

To properly exclude \textit{unphysical} degrees of freedom, it proves advantageous to localise the latter on distinguished \textit{mathematical} degrees of freedom.
This can be achieved in unitary gauge where the physical subspace $\mathcal{H}_{\mathbb{Z}_2GH}=\{\Ket{\Psi}\,|\,G_s=\mathds{1}\}$ is unitarily rotated into
the new subspace $\tilde{\mathcal{H}}_{\mathbb{Z}_2GH}=\{\Ket{\Psi}\,|\,\sigma_s^x=\mathds{1}\}=T\mathcal{H}_{\mathbb{Z}_2GH}$.
The hermitian and unitary transformation reads
\begin{equation}
  T=\prod_{e\in\mathbb{E}}\left[\mathds{1}_e P_e^++\tilde I_e P_e^-\right]
\end{equation}
with the projectors $P_e^\pm=\frac{1}{2}(\mathds{1}_e\pm\tau_e^x)$ and the operator $\tilde I_{e=(st)}\equiv \sigma_s^z\sigma_t^z$.
To transform the jump operators, it is useful to show first that
\begin{subequations}
  \begin{eqnarray*}
    T\tau_{e}^z T^\dag&=&I_e,\quad T\tau_e^x T^\dag=\tau_e^x\\
    T\sigma_s^z T^\dag&=& \sigma_s^z,\quad T\sigma_s^x T^\dag=G_s
  \end{eqnarray*}
\end{subequations}
and then calculate $\tilde L=T L T^\dag$; this yields the unitary gauge representation in the right-hand column of Tab.~\ref{tab:LGT_setup_large}.
The gauge condition becomes trivial, $\sigma_s^x=\mathds{1}$, and can be accounted for by just dropping the matter field completely as it does not
enter the dynamics. This establishes a one-to-one correspondence between mathematical and physical degrees of freedom which prevents the mean field
theory from taking into account the unphysical ones. To this end, we make the ansatz
$\rho=\bigotimes_{e\in\mathbb{E}}\rho^{\text{g}}_e$ and derive once again the (now three dimensional) dynamical mean field flow $\vec F(\vec g)$.

The results are illustrated in Fig.~\ref{fig:LGT_plots_B}.
In accordance with the mean field theory for the Hamiltonian counterpart, the distinction between Higgs and free charge phase is lost whereas
the analytical path between confined charge and Higgs phase is recovered, see (A). The phase transition separating confined charge and free charge \& Higgs phase 
is still discontinuous as (B) reveals. In contrast to the mean field approach discussed in the main text, the flow $\vec F(\vec g)$ is only three dimensional and we
can illustrate its topology faithfully in the Bloch ball cross sections (C). Note that the $g_z>0$ solution is indeed \textit{stable} since there is
a nearby unstable fixed point separating the stable $g_z>0$ and $g_z=0$ solutions. The cross sections illustrate nicely how the topology
of the mean field flow gives rise to the continuous transition connecting the two phases: When the discontinuous phase boundary is traversed from (4) to (3),
the $g_z=0$ solution remains at the center of the Bloch ball while close to the $g_z=1$ pole two new fixed points (one stable, one unstable) emerge.
When, in contrast, the continuous path along (2) is taken, the single stable solution approaches the pole until it ``splits'' into a pair of stable and an unstable
fixed point; one stable fixed point approaches the pole while the other seeks the center of the Bloch ball.


\begin{acknowledgments}
  \paragraph{Acknowledgements.}
  We acknowledge support by the Center for Integrated Quantum Science and Technology (IQST) 
  and the Deutsche Forschungsgemeinschaft (DFG) within SFB TRR 21. 
  We thank J. K. Pachos for inspiring discussions.
  NL thanks the German National Academic Foundation for their support.
\end{acknowledgments}





\end{document}